%% file: boson_crystal6cm.tex
\documentclass[prb,aps,twocolumn,showpacs,preprintnumbers,floats,amsmath,amssymb]{revtex4}

\usepackage{graphicx}
\usepackage{dcolumn}
\usepackage{bm}

\def \be{\begin{equation}}
\def \ee{\end{equation}}

\def\moy#1{\left\langle #1 \right\rangle}

\begin{document}

\title{Quantum melting of a crystal of dipolar bosons}

\author{Christophe Mora$^{1,2}$}
\affiliation{$^1$~Laboratoire Pierre Aigrain, D\'epartement de Physique,
Ecole Normale Sup\'erieure,
24 rue Lhomond, F-75005 Paris, France}
\affiliation{$^2$~Institut f\"ur Theoretische Physik,
Heinrich-Heine Universit\"at, D-40225 D\"usseldorf}
\author{Olivier Parcollet$^3$}
\affiliation{$^3$ Service de Physique Th\'eorique, CEA/DSM/SPhT-CNRS/SPM/URA 2306
CEA Saclay, F-91191 Gif-Sur-Yvette, France}
\author{Xavier Waintal$^4$}
\affiliation{$^4$~Nanoelectronics group, Service de Physique de l'\'Etat Condens{\'e},
CEA Saclay, 91191 Gif-sur-Yvette cedex, France}

\date{\today}

\begin{abstract}
We investigate the behaviour of dipolar bosons in two dimensions. We describe the large density
crystalline limit analytically while we use quantum Monte-Carlo to study the melting toward
the Bose-Einstein condensate. We find strong evidence for a first order transition.
We characterize the window of experimentally accessible parameters in the context of
ultracold bosons and show that observing the quantum melting should be within grasp
once one is able to form cold heteronuclear molecules.
Close to the melting, we can not conclude on the existence
of a supersolid phase due to an insufficient overlap of our variational Bijl-Jastrow Ansatz
with the actual ground state.
\end{abstract}

\maketitle

\section{Introduction}
Quantum melting (the melting of a crystal induced by quantum fluctuations) can
be very different from a classical melting (induced by temperature). In two dimensional
electronic systems, it was found that near the melting point~\cite{tanatar1989}, the system is neither in
a liquid nor in a solid phase but shares properties with both phases.~\cite{waintal2006,falakshahi2005} 
In bosonic systems,
such an intermediate phase ("supersolid" between the solid and the superfluid phase) was 
proposed many years ago.~\cite{andreev1969,leggett1970,chester1970} The supersolid phase can be thought as a condensate of
either static or dynamical defects.~\cite{prokofev2005} Some supersolid behaviour (i.e. existence of
a non zero superfluid fraction in a solid system) has indeed been recently observed~\cite{kim2004}
in Helium 4, creating a large renew of interest on both theoretical and experimental sides.
The situation in Helium 4 is however controversial as (i) uptodate quantum Monte-Carlo simulations
do not find any superfluid fraction in the solid phase\cite{clark2006,ceperley2004,boninsegni2006} while
providing very precise predictions in the superfluid and the solid phase, (ii) the experiments seem to be 
sensitive to the way the solid is prepared~\cite{rittner2006} (an annealing of the supersolid kills or enhances 
the superfluid fraction depending on the experiments) indicating that disorder probably
plays a key role in this physics and (iii) one gets only access experimentally to macroscopic 
quantities which make it difficult to infer the physical mechanisms actually involved.

An alternative bosonic system is now promising due to recent progresses in the field
of ultracold gases. The production and observation of heteronuclear alkali molecules~\cite{doyle}
at ultralow temperatures 
has been reached with various dimers, {\it e.g.} KRb~\cite{wang2004,mancini2004}, 
RbCs~\cite{sage2005} or LiCs~\cite{kraft2006}, yet not in their true 
molecular ground state. Polar molecules confined in a two-dimensional trap
with a perpendicular electric field could be given an important dipole moment
(on the order of $1$ Debye~\cite{aymar2005}) leading to strong repulsive dipole-dipole interaction 
on the plane. Although these interactions are not trully long range in $2$D, the 
potential range is larger than for typical van der Waals interaction
and smaller densities are necessary to observe crystallization.
Moreover the true repulsive character of dipole-dipole interaction should strongly inhibit
inelastic three-body recombination. Such  systems
could be  very good candidates for the study of quantum melting as they do not share the
usual limitations found in Helium 4: there is a complete control of the external potential seen by
molecules (no disorder) and many direct observations can been done (of the spatial distribution
of the density for the solid, of the condensate fraction for the superfluid part).
It is the purpose of this paper to study this potential use of cold  dipolar gases in some 
detail. We note that two papers with a similar topic and some overlapping results
appeared recently~\cite{buchler2007,astrakharchik2007} and hence, whenever appropriate 
we will compare our findings with these works. Another possible experimental realization of the model 
studied in this paper are electron-hole heterostructure bilayers where the excitons also interact through dipole-dipole
interaction.~\cite{voros2005} 

After introducing the model (Bosons in two dimensions with a dipole-dipole repulsion),
and briefly the numerical technique (Green Function Monte-Carlo or GFMC) in section~\ref{sec:model},
 we proceed with the description of the high density limit
where the crystal is expected. In section~\ref{sec:phonons} we calculate the phonon spectrum of the
crystal, from which we extract the high density expansion of the energy of the system.
The nature (first order) of the transition is then discussed in section~\ref{sec:nature} using the
GFMC technique. Particular attention is paid to the sensitivity of the results to the trial wave-function used
in the calculation. The quality of the wave-function is usually measured in term of the distance in energy
to the true ground state. We find that a better measure can be done by calculating the overlap of the trial
wave-function with the true ground-state~\cite{mora2007}. 
In section~\ref{sec:supersolid}, we discuss the possibility of the existence
of a supersolid in this system. In the last section~\ref{sec:conclusions}, we estimate the regime of parameters
that should be accessible experimentally.
The appendix contains the Ewald summation technique that is applied to the dipole-dipole interaction
to reduce finite size effects.

\section{The Model: $N$ bosons in 2 dimensions with dipole-dipole interaction}
\label{sec:model}

Our system is made of $N$ bosons of mass $M_0$
confined in two dimensions which interact
through a dipole-dipole interaction of strength $C_{\rm dd}$,
so that the Hamiltonian reads
\begin{equation}\label{hamilto1}
H = - \frac{\hbar^2}{2 M_0} \sum_{i=1}^{N} \nabla_i^2 + C_{\rm dd}
\sum_{i<j} \frac{1}{|{\bf r}_i - {\bf r}_j|^3},
\end{equation}
with dipoles polarized perpendicular to the plane.
Measuring lengths in unit of the average distance $\ell$
between bosons (defined as  $\pi\ell^2\equiv 1/n$ where $n$ is the $2$D 
density), and the energies in unit of $E_B\equiv C_{\rm dd}/(2\ell^3)$,
the rescaled Hamiltonian depends on a unique dimensionless parameter $r_s$
and reads,
\begin{equation}\label{hamilto2}
H = - \frac{1}{r_s} \sum_{i=1}^{N} \nabla_i^2 + 2 \sum_{i<j}
\frac{1}{|{\bf r}_i - {\bf r}_j|^3}.
\end{equation}
The parameter $r_s$ (named in analogy with electronic systems)
roughly measures the ratio between interaction energy
$\sim C_{\rm dd}/\ell^3$ and kinetic energy $\sim \hbar^2/(M_0 \ell^2)$
and reads,
\begin{equation}
r_s \equiv \ell_B^* / \ell,
\end{equation}
where $\ell_B^* = M_0 C_{\rm dd} / \hbar^2$. Contrary to electronic systems, 
$r_s$ increases with density, so that Bose-Einstein condensation is obtained
at low density while at high density the dipole-dipole interaction dominates
and the system crystallizes into a triangular lattice (see Sec.~\ref{sec:phonons}). 

\subsection{The Green Function Monte-Carlo technique}
The ground state properties of the Hamiltonian Eq.~\eqref{hamilto2}
can be conveniently studied using quantum Monte-Carlo techniques which for a bosonic
system (in contrast to fermionic systems where one has to deal with the so called sign problem)
provides access to the exact ground state $|\Psi_0\rangle$ of the system. The idea behind this
method is to project an initial (variational) Guiding Wave Function $|\Psi_G\rangle$ (GWF) on the 
exact ground state $|\Psi_0\rangle$ by applying
the projection operator $e^{-\beta H}$. For large projecting time $\beta$, the excited states are 
exponentially filtered out from the initial trial wave-function, and one gets,
\be
\label{gfmc}
|\Psi_0\rangle \propto \lim_{\beta\rightarrow\infty} e^{-\beta H}|\Psi_G\rangle
\ee
GFMC is a scheme to implement Eq.~(\ref{gfmc}) in a stochastic way. In practice, $|\Psi_G\rangle$  not only 
enters as the starting point
but also through the way the Hilbert space is sampled when the projection operator is applied. A good choice of 
the trial wave-function is therefore important to get a good signal to noise ratio. In
principle however (i.e. for infinite computing time), the results should be independent of $|\Psi_G\rangle$ . 

Although this technique can be used for continuum systems, our particular implementation
(which has been described in Ref.~\onlinecite{waintal2006}) relies on the discretization of the Hamiltonian on a grid
(which in returns allows us to work without discretizing the imaginary time $\beta$). For some calculations,
we have also used the recently introduced Reptation Quantum Monte Carlo algorithm 
(RQMC)~\cite{baroni1999,pierleoni2005}. RQMC is formally 
equivalent to GFMC and is its path integral counter part. For the regimes
discussed in this paper, we found however that RQMC gave no real advantage over GFMC. Unless explicitly stated
we use only pure estimators (forward walking technique~\cite{sorella1998}) for the quantum averages.

\subsection{Discretized Hamiltonian}
The bosonic system is put on a rectangular
grid of $L_x \times L_y$ sites with nearest neighbor hopping.
Finite size effects are considerably reduced by repeating periodically
the simulation box on the two-dimensional plane. This amounts to use periodic
boundary conditions for hopping and to replace the dipole interaction
by an effective two-body interaction that includes all periodic
images. The resulting summation is efficiently performed using a straightforward
extension of the Ewald summation technique. This is discussed in more details
in appendix~\ref{appen:ewald}.
The discretized Hamiltonian on the grid reads
\be
\label{eq:model}
H=-t\sum_{\langle\vec r,\vec r'\rangle}c_{\vec r}^\dagger c_{\vec r'}
+\frac{U}{2} \sum_{\vec r\ne\vec r'} V(\vec r-\vec r\,') n_{\vec r} n_{\vec r'}  + (4t+\lambda) N ,
\ee
where the operator $c_{\vec r}^\dagger$ ($c_{\vec r}$) creates (destroys) a boson on
point $\vec r$ with the standard commutation relation rules, the sum $\sum_{\langle\vec r,\vec r'\rangle}$
is done on the nearest neighbor points on the grid.
The form of the periodized two-body interaction $V(\vec r)$ and of the constant $\lambda$ (which accounts for the 
interaction of a boson with its own images) are given in appendix~\ref{appen:ewald}. 
Noting $\nu=N/(L_x L_y)$ the filling factor, the continuous model
Eq.(\ref{hamilto2}) is reproduced by choosing the hopping amplitude to be $t = 1/ (r_s \pi \nu)$  and the 
effective interaction strength $U=2/(\pi \nu)^{3/2}$ in the limit $\nu \to0$ . In practice we took
$\nu=1/56$ (some small influence of the presence of the grid) down to $\nu=1/780$ (no detectable influence of the grid).
The grid dimensions are also chosen in order not to induce distortion of the triangular lattice of the crystal.

\subsection{The guiding wave-functions}
All our Monte Carlo simulations start from a trial wavefunction which is an initial guess for
the exact ground state wavefunction. 
The trial GWF is also used as
a guide to converge to the ground state during the GFMC projection.
For practical efficiency, {\it i.e.} to reduce
noise in numerical results, it is important
that the GWF is as close as possible from the unknown exact solution.
We use here a Bijl-Jastrow form: 
\begin{equation}\label{guiding}
\Psi_{GWF} ({\bf r}_1,\ldots,{\bf r}_N) = 
\prod_{i=1}^N \phi_1({\bf r}_i) \prod_{j<k} \phi_2 (|{\bf r}_j - {\bf r}_k|)
\end{equation}
which includes at the variational level one-body and two-body correlations.
Two-body correlations are essential in the homogeneous (BEC) limit where it dominates the physics
and determines the shape of the wavefunction.
On the contrary, one-body terms are sufficient to describe the crystal limit
where atoms are almost frozen. 
It is worth noticing that the two-body scattering problem can be solved exactly at
zero energy, leading $\phi_2^{\rm t-b} (r) = A K_0 (2 \sqrt{r_s/r})$, where $A$ is an arbitrary constant
and $K_0$ the modified Bessel function of the second kind. Its short distance behaviour is
 $\phi_2^{\rm t-b} (r) \propto (r/r_s)^{1/4} e^{-2 \sqrt{r_s/r}}$. In practice, we chose
\be
\phi_2 (r) = \exp \left( -2 \sqrt{\frac{r_s}{r}} e^{-r/A} \right),
\ee
with a single variational parameter $A$ which sets the length over which two-body correlations
are suppressed. In particular, it reproduces the leading short distance behavior for the
two-body scattering problem. More complicated variational forms have been tested (with more
variational parameters) with the same efficiency.
Typical values of $A\simeq1$ have been found in the vicinity of the quantum melting 
point $r_s \simeq 27$.

Noting $\Delta y$ ($\Delta x$) the distance between sites along the $y$ ($x$) axis ($\Delta y/ \Delta x
= \sqrt{3}/2$), we define the 
vector ${\bf q}_1 = (0,2 \pi / \Delta y)$ in the reciprocal lattice. Vectors ${\bf q}_{2/3}
= (\pm 2 \pi / \Delta x,-\pi/\Delta y)$
are obtained by rotating ${\bf q}_1$ by an angle of $2 \pi /3$ and $-2\pi/3$.
We choose for the one-body trial function
\be\label{trial}
\phi_1 ( {\bf r} ) = \prod_{i=1,2,3} \left ( 1+ \alpha \cos ( {\bf q}_i \cdot {\bf r} ) \right),
\ee
where maxima reproduce the triangular lattice expected for the crystal.
This function is well-suited to describe quantum melting since it interpolates between a flat
liquid (BEC)-type pattern for $\alpha=0$ and a triangular crystal form for $\alpha \ne0$.
Moreover it conserves the symmetrization property imposed by bosonic statistics 
in contrast with the Gaussian form often considered in the literature.

\section{Properties of the crystal phase}
\label{sec:phonons}
The infinite $r_s$ limit corresponds to a purely classical system 
of dipoles with interaction energy only. We have compared energies of various lattice configurations
and found that
the ground state is given by the triangular Bravais lattice of energy per particle
$E_0 = 1.59702 \ldots$.
For comparison, the square lattice energy is $E_0=1.62232\ldots$.   
Ground state properties of the crystal phase can be computed perturbatively for large $r_s$
following Ref.\cite{maradudin1977}. For large but finite $r_s$, kinetic energy plays a role as 
atoms are able to move around the perfect triangular lattice configuration. We treat kinetic terms perturbatively
which amounts to an harmonic approximation for the potential felt by the atoms.
The emerging collective modes are then a sum of harmonic oscillators over the Brillouin zone.
They describe phonons and dominate dynamical properties of the ground state.
Moreover first order correction in $1/\sqrt{r_s}$ to the ground state energy derives from 
zero-point energies of these vibrational modes.

\begin{figure}
\includegraphics[width=8cm]{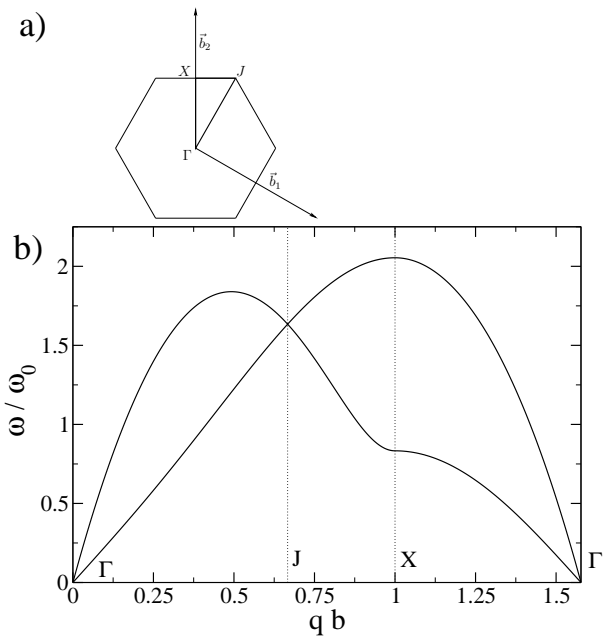}
\caption{\label{fig:phonons} $a)$ First Brillouin zone for a triangular lattice. The corresponding 
irreducible zone is delimited by the points $\Gamma$, $J$ and $X$.
$b)$
Dispersion curves for phonon modes in the large $r_s$ limit
(crystal). Each branch corresponds to a particular polarization. The wavevector ${\bf q}$ is taken
on the boundary of the irreducible Brillouin zone along the path 
$\Gamma \to J \to X \to \Gamma$.}
\end{figure}

The Hamiltonian Eq.~\eqref{hamilto2} expands around the classical stable position
${\bf r}_i = {\bf R}_i + {\bf u}_i$ where $\{{\bf R}_i\}$ denotes the triangular 
lattice configuration. Stopping at second order in atomic displacements, 
it is straightforward to diagonalize the Hamiltonian in Fourier space. Collective eigenmodes read
${\bf u}_j (t) = {\bf \bar{u}}_{\bf q} e^{i ( {\bf q}\cdot {\bf R}_j - \omega t)}$ where 
the vectors ${\bf \bar{u}}_{\bf q}$ are solutions of the eigenvalue problem
\begin{equation}\label{eigenproblem}
\omega^2({\bf q}) {\bf \bar{u}}_{\bf q} = \frac{4}{r_s} \Phi ({\bf q}) {\bf \bar{u}}_{\bf q}.
\end{equation}
Here $\Phi  ({\bf q})$ is a tensor or $2\times2$ matrix defined by
\begin{equation}\label{kernel}
\Phi_{\alpha,\beta}  ({\bf q}) = \left. \frac{\partial^2}{\partial u_\alpha \partial u_\beta} 
\left( \sum_{j\ne0} \frac{1}{|{\bf R}_j + {\bf u}|^3} \left( 1- e^{i {\bf q}\cdot {\bf R}_j} \right)
\right) \right|_{{\bf u} \to0},
\end{equation}
with Cartesian coordinates $\alpha,\beta=x,y$. For each wavevector ${\bf q}$ in the first Brillouin
zone, Eq.~\eqref{eigenproblem} has two solutions which leads to two branches of collective excitations.
The symmetry of the triangular lattice under the point group $C_{\rm 6v}$ allows to restrict 
values of  ${\bf q}$ within the irreducible Brillouin zone represented in Fig.\ref{fig:phonons}.
For illustration we show in  Fig.~\ref{fig:phonons} the corresponding 
phonon eigenmodes, solutions of Eq.~\eqref{eigenproblem} ,
along the boundary of the irreducible Brillouin zone.
We rescale  Eq.~\eqref{eigenproblem} with the definition
\begin{equation}
\omega_0 = \frac{8}{b^{5/2} \, \sqrt{r_s}},
\end{equation}
where $b=\sqrt{2/\sqrt{3}}$ is the lattice spacing for the triangular lattice with density $1/\pi$.
$\bar{\omega}^2 = ( \omega/\omega_0)^2$ is now eigenvalue of $\bar{\Phi} = (b^5/16) \Phi$. 
For practical purposes, the kernel $\bar{\Phi} ({\bf q})$ is not computed by direct summation 
but we use a variant of the Ewald summation technique instead, see appendix~\ref{appen:ewald}.

Correction to the ground state energy with respect to the classical case ($r_s \to +\infty$)
stems perturbatively from the zero-point energies of these phononic modes. Explicitly, the 
ground state energy per particle is given by
\[
E = \sum_{i \ne0} \frac{1}{|{\rm R}_i|^3} + \sum_{j=1,2} \int_{BZ} 
\frac{d^2 q}{V_{BZ}} \, \frac{\omega_j ({\bf q}) }{2},
\]
where $j=1,2$ stands for the two phonon branches and $V_{BZ} = 8 \pi^2 /\sqrt{3} b^2$ is the
area of the Brillouin zone (BZ). In reduced units it takes the form
\begin{equation}
E = E_0 + \frac{1}{\sqrt{r_s}} \left( \frac{24\sqrt{3}}{b^{5/2}} \right)
\sum_{j=1,2} \int_{IBZ} d^2 q  \, \left( \frac{b}{2\pi} \right)^2 \bar{\omega}_j ({\bf q}),
\end{equation}
where the summation is performed over the irreducible Brillouin zone (IBZ), see Fig.~\ref{fig:phonons}a)).
The numerical evaluation gives
\begin{equation}\label{expansion-wigner}
E = 1.59702 + \frac{2.02987}{\sqrt{r_s}} + \mathcal{O} \left( \frac{1}{r_s} \right).
\end{equation}
At a not too large value of $r_s=50$ and for $N=50$ particles the GFMC simulation gives
 $E_{\rm GFMC}=1.893$ while the aforementioned expansion provides $E = 1.884$, {\it i.e.}
within $0.5 \%$ of the exact result.

The ground state wavefunction  is a simple product of all 
Gaussian ground states of the phonon modes. It can be written as
\[
\Phi_{\rm G} ( \{ {\bf u}_i \}) = \mathcal{N} e^{- \sum_{j=1,2} \int_{BZ} \frac{d^2 q}{V_{BZ}} \, 
{\bf \bar{u}}_{{\bf q},j}  {\bf \bar{u}}_{-{\bf q},j} r_s \omega_{j} ({\bf q}) /4}  ,
\]
with $\mathcal{N}$ a normalization factor. Vectors ${\bf \bar{u}}_{\bf q}$
are Fourier transforms of the original variables $\{ {\bf u}_i \}$ and eigenvectors of the
system~\eqref{eigenproblem}. The reduced onebody wavefunction is obtained by integrating all
atomic positions but one. This leads to the following anisotropic Gaussian wavefunction
\[
\Phi_{{\rm G},1}(x,y) = \mathcal{N}_1 e^{ -  [ (x/x_0)^2 + (y/y_0)^2 ]/2},
\]
with typical sizes $x_0$ and $y_0$ respectively in the $x$ and $y$ directions.
Let $(\alpha ({\bf q}), \beta ({\bf q}))$ and  $( -\beta ({\bf q}),\alpha ({\bf q}))$ be the
normalized ($\alpha ({\bf q})^2 + \beta ({\bf q})^2=1)$ eigenvectors of Eq.~\eqref{eigenproblem},
we find
\[
\sqrt{r_s} x_0^2 =  \frac{3\sqrt{3}}{2} b^{9/2} 
 \int_{IBZ} \frac{d^2 q}{(2 \pi)^2}  \,  
\left( \frac{\alpha({\bf q})^2}{\bar{\omega}_1 ({\bf q})} 
+ \frac{\beta({\bf q})^2}{ \bar{\omega}_2 ({\bf q}) } \right),
\]
and the same expression holds for $y_0$ where $\alpha ({\bf q})$ and $\beta ({\bf q})$
are interchanged. A numerical evaluation yields $x_0 = 1.3613 /r_s^{1/4}$ and
$y_0 = 0.9509/r_s^{1/4}$ in units of $\ell$.

\section{First order nature of the quantum melting}
\label{sec:nature}
In this section, we use the quantum Monte-Carlo calculations in order to determine the order
of the transition. We find strong numerical evidences of a first order transition. 

Close to a first order transition, the GFMC energies are expected to depend on the GWF $|\Psi_G\rangle$.
To illustrate this point qualitatively, let us expand $|\Psi_G\rangle$ on the eigenbasis
of the Hamiltonian,
\be
|\Psi_G (\alpha) \rangle \propto |\Psi_{\rm BEC} \rangle + c(\alpha )|\Psi_{\rm cry}\rangle  +\cdots
\ee
where $|\Psi_{\rm BEC}\rangle$ ($|\Psi_{\rm cry}\rangle$) stands for the BEC (crystal) eigenstate and
$c(\alpha)$ measures the relative projection of the GWF on the two competing eigenstates (a more careful
treatment would also include the low lying excitations of the two phases). After projection, 
one gets a state $|\Psi\rangle = e^{-\beta H}|\Psi_G (\alpha) \rangle$,  
\be
\label{eq:firstorder}
|\Psi\rangle \propto |\Psi_{\rm BEC} \rangle + c(\alpha )e^{-\beta (E_{
\rm BEC}-E_{\rm cry})} |\Psi_{\rm cry}\rangle  
\ee  
where $E_{\rm BEC}$ ($E_{\rm cry}$) stands for the BEC (crystal) energy. 
For very large imaginary time $\beta$, the exponential is very large (or small), and $|\Psi\rangle$
converges to whichever of the BEC or crystal state is more stable. Real simulations however are
done with finite values of $\beta$ (typically $30/t$ in our case). As one approaches the transition, 
$E_{\rm BEC}-E_{\rm cry}$ vanishes, hence the exponential term tends to one, and one is left with a (incoherent)
superposition $|\Psi\rangle \propto |\Psi_{\rm BEC} \rangle + c(\alpha )|\Psi_{\rm cry}\rangle$
whose properties are sensitive to the choice of the variational parameter $\alpha$ of the GWF.  

\begin{figure}
\includegraphics[width=8cm]{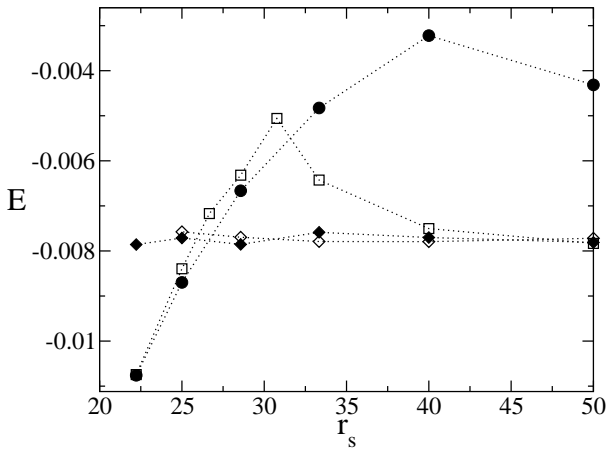}
\caption{\label{fig:crossing} GFMC calculation of the energy per particle for various values of the symmetry
breaking parameter $\alpha=0$ (circles), $\alpha=0.2$ (empty squares), $\alpha=0.65$ (diamonds) and $\alpha=0.75$
(empty diamonds). The transition is emphasized by plotting $E_{\rm GFMC} - 1.59702 - 2.02987/\sqrt{r_s} 
- 0.85/r_s$ as a function of $r_s$ (the last number here, $0.85$, is a fitted parameter). 
Above the critical $r_s\approx27\pm 1$, the
symmetry breaking state is energetically favored. This plots also demonstrates that the transition
is first order. The system contains $N=72$ bosons in $48\times 84$ sites. }
\end{figure}

In Fig.\ref{fig:crossing}, we present the GFMC energies as a function of $r_s$ for various choices
of the symmetry breaking parameter $\alpha$. Those data support the previous picture: concentrating at first on
the $\alpha=0$ ("BEC" like state, with $c(0)\ll 1$) and $\alpha=0.75$ ("crystal" like state, with $c(0.75)\gg 1$)
curves, we find a crossing around $r_s\approx 27$ above which the crystal becomes more stable than the BEC. When the
symmetry is only weakly broken ($\alpha=0.2$), the GFMC result is closed to the BEC state close
to the transition. As one increases $r_s$ further, the energy difference $E_{\rm BEC}-E_{\rm cry}$ increases and
the exponential in Eq.~(\ref{eq:firstorder}) eventually compensates the
smallness of $c(0.2)$ and $|\Psi\rangle $ converges toward the actual ground state $|\Psi_{\rm cry}\rangle$.

\begin{figure}
\vglue +0.05cm
\includegraphics[width=8cm]{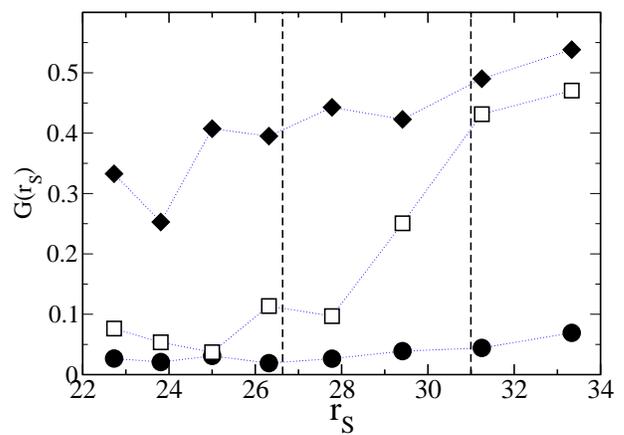}
\caption{\label{fig:order} Crystal order parameter $G(r_s)$ for three value of the symmetry breaking 
parameter $\alpha=0$ (circles), $\alpha=0.4$ (empty squares) and $\alpha=0.8$ (diamonds). The system 
contains $N=72$ bosons in $48\times 84$ sites and $G(r_s)$ has been evaluated at $\beta= 30$. The vertical 
dashed line indicate the region where the results are very sensitive to the choice of the wave-function.}
\end{figure}

\begin{figure}
\vglue +0.05cm
\includegraphics[width=8cm]{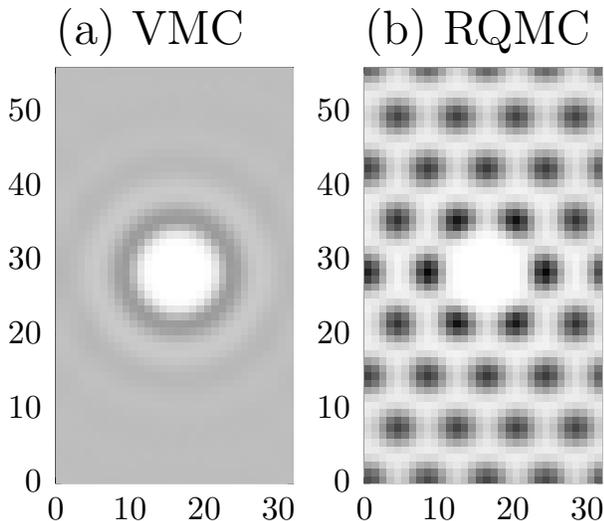}
\caption{\label{fig:VMC_vs_RQMC} Density-density correlation function $g(\vec r)$ at
$r_s=50$, well above the melting point. The calculations are done for $32$ particles in a $32\times 56$ grid
with the trial wave-function for the condensate, $\alpha=0$. On the left, the VMC results (a) show no sign of broken symmetry while after projection on 
the ground state, the RQMC results (b) for $\beta=20$ are in a crystal state. The gray scale goes from zero (white) to $4$ (black).  }
\end{figure}

To further characterize the transition, we now look at the crystalline order parameter. Introducing the static
density-density correlation function $g(\vec r)$ (that roughly measures the probability of finding a particle on
point $\vec r$ knowing that there is one at point $\vec 0$),
\be
g(\vec r)=\frac{L_xL_y}{N(N+1)}\sum_{\vec h}\moy{c_{\vec r+\vec h}^\dagger c_{\vec h}^\dagger c_{\vec h} c_{\vec r+\vec h}}
\ee
the order parameter is usually defined as the Fourier transform of $g(\vec r)$ taken at a $\vec k$ vector belonging
to the reciprocal lattice of the expected crystal. A typical example of $g(\vec r)$ in the BEC (crystal) case can
be found in Fig.~\ref{fig:VMC_vs_RQMC}a  (Fig.~\ref{fig:VMC_vs_RQMC}b). Here we define the crystalline order parameter 
$G(r_s)$ (in an equivalent but more convenient way) as the average value of $g(\vec r)$ at the peaks corresponding to the 
classical position of the crystal (minus its average value). In Fig.\ref{fig:order}, we plot $G(r_s)$ for non broken ($\alpha=0$),
broken ($\alpha=0.8$) and partially broken ($\alpha=0.4$) symmetry. The non broken and broken symmetry case have 
respectively zero and non zero $G$, hence $G$ is discontinuous at the transition which should therefore be first order.
The case with partially broken symmetry shows the expected crossover near the transition point.

Far from the transition point, the results are supposed to be independent of the GWF. We checked that this is indeed
the case in our calculation: Fig.~\ref{fig:VMC_vs_RQMC} presents the results before (Variational Monte-Carlo VMC, 
on the left) and after (RQMC, on the right) projection on the ground state. One can see that although the starting 
point has no broken symmetry ($\alpha=0$), the RQMC (or GFMC) algorithm is able to converge to the 
crystal ground state nevertheless.
This result was obtained for a rather small system of $N=32$ particles however, and for a larger system of $N=72$ bosons,
a small amount of $\alpha$ was needed in order to get the correct ground state.

The above results are consistent with those obtained in Ref.~\cite{buchler2007,astrakharchik2007}. 
In Ref.~\onlinecite{buchler2007},
the authors use a finite temperature algorithm and find the transition at $r_s = 32\pm 7$. Indeed, in our unit,
they work at a temperature of the order of $5.\ \ 10^{-3}$, i.e. several time the difference of energy (per
particle) between the BEC and the crystal phase near the transition, see fig.~\ref{fig:VMC_vs_RQMC}. 
These explains the relatively important error
bar on their critical value of $r_s$ and the fact that their superfluid fraction switches back and forth between 0
and 1 close to the transition. In Ref.~\cite{astrakharchik2007}, the authors use the DMC technique, 
similar to ours. 
The GWF that they are using to describe the crystal is not symmetric with respect to the interchange of the 
bosons, and hence describe a crystal of distinguishable particles. However they find a critical $r_s=30$
which is consistent with ours (we found upon decreasing the filling factor $\nu$ a small increase of the
critical value of $r_s$).

We now proceed with the Maxwell construction, in order to find the region where
phase separation is to be expected. In the rest of this section, we measure energies
in unit of $\bar E_B= C_{dd}/(2a_B^3)$ and not $E_B$ as in the rest of the paper since
$E_B$ depends on $r_s$.
Defining the density $\bar n=r_s^2$, such a phase separation
can occur between the two values of the density $\bar n_1$ and $\bar n_2$ satisfying 
$[E(\bar n_1)-E(\bar n_2)]/(\bar n_1-\bar n_2)= \partial E/\partial \bar n (\bar n_1)=
\partial E/\partial \bar n (\bar n_2)$. Close to the transition, we find that a good
parameterization of the energy reads,
\be
E\approx a_0 \bar n^{3/2}+ a_{1/2} \bar n^{5/4} +a_1 \bar n^{} + \eta \theta(\sqrt{\bar n}-r_s^* )
\left[\sqrt{\bar n} - r_s^*  \right] \bar n^{3/2}
\ee
where $a_0\approx 1.592$, $a_{1/2}\approx 2.03$, $a_1\approx 0.85$, $\eta\approx 5\  10^{-4}$, $r_s^*\approx 27$,
 and $\theta(x)$ is the Heaviside function discriminating the two phases. Note that
the coefficient $\eta$ can be directly extracted from Fig.~\ref{fig:crossing}.
We find that the region where phase separation can occur is very small, given by 
a width $\delta r_s=\alpha (r_s^*)^2/(3 a_0)$ around the transition point $r_s^*$.
For our parameters, it translates into $\delta r_s \approx 0.04$ which is very small.

\section{On the presence of a supersolid phase}
\label{sec:supersolid}
Once the presence of a first order transition has been established, a natural question
is whether the superfluid fraction vanishes right at the transition or if there is a region
where both the superfluid fraction and the order parameter $G$ are non zero simultaneously. 
Such a region would be a supersolid phase and has been searched for for many years. However,
to our knowledge, the only model for which it has actually been found are bosons on a 
lattice~\cite{otterlo1994,batrouni1995,goral2002}
at large filling factors (i.e. systems similar to our but in the limit $\nu\approx 1$ not $\nu\ll 1$) 
where the crystal is somehow stabilized by the underlying grid. When the crystal is uniquely due
to the repulsive interaction (as in Helium 4 or in the system considered here), no convincing evidence could be
gathered so far.

\begin{figure}
\vglue +0.05cm
\includegraphics[width=8cm]{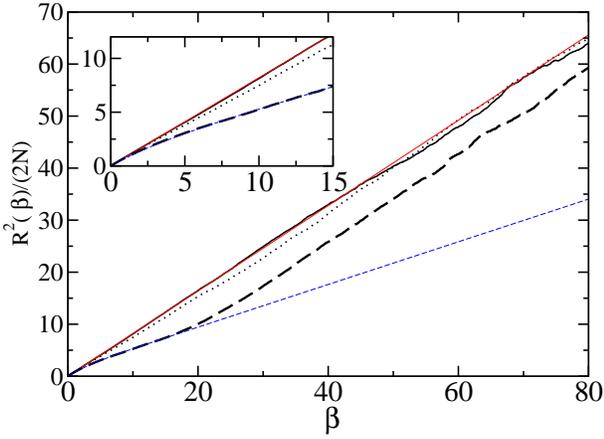}
\caption{\label{fig:super} (color online). Diffusion of the center of mass $R^2(\beta)$ for $N=8$ particles in $16\times 28$ sites
at $r_s=200$ using $10^6$ walkers. The various lines correspond to $\alpha=0$ (straight),
$\alpha=0.2$ (dotted) and $\alpha=0.5$ (dashed). The thin red (blue dashed) lines are fit $y=0.82 \beta$
( $y=0.41\beta + 1.24 (1-e^{-\beta/3.4})$ ) corresponding to a superfluid fraction of $0.82$ and
$0.41$ respectively. Inset: zoom of the main figure.}
\end{figure}

There are two alternative way to probe the superfluid character of the system. The direct way consists in
calculating directly the superfluid fraction $\rho_S$ of the system. It is defined as the fraction of the bosons that 
does not follow when the frame is put (adiabatically) into motion (rotation). It is the bosonic equivalent
of the curvature of the energy with respect to an Aharonov flux for electrons. For quantum Monte-Carlo calculations, it is 
can be conveniently calculated using the diffusive motion of the center of mass of the system (in imaginary
time) as
\be
\rho_S =\frac{1}{2N}\lim_{\beta\gg 1}\frac{R^2(\beta)}{\beta}
\ee
where $\beta$ is the imaginary time and $R^2(\beta)$ is the second moment of the center of mass 
in one direction (say $x$). Such a formula can be easily implemented, but the results should be
taken with care. For $r_s\le 27$ (BEC state) we find that $\rho_S\approx 1$. For large $r_s$, we find
that $\rho_S\approx 0$. However, for $r_s\ge 27$ but close to the transition, we could not get a reliable
measure of  $\rho_S$ and hence cannot conclude about the presence of a supersolid in this model. This is due to a combination
of two difficulties: (i) close to the transition, "metastable state" can enter into the GWF, hence artificially
enhancing $\rho_S$ (depending on the choice of $\beta$, see the discussion in section~\ref{sec:nature}),
(ii) In the crystal phase, the superfluid fraction is given by rather rare events that 
contain "dynamical defects"~\cite{prokofev2005}.
To illustrate the difficulty of the calculation, let us look at a small system of $N=8$ particles,
where using brute force simulations, reliable results could be obtained for any choice of $\alpha$. Using a very
large number of walkers ($10^6$), and rather long simulations ($\beta=80/t$), we find in Fig.\ref{fig:super} a
superfluid fraction $\rho_S\approx 0.82$ for all values of $\alpha$. However, for $\alpha=0.5$ the correct superfluid
fraction is only recovered after a (large) simulation time ($\beta\ge 20$). A careful analysis shows that for 
$\beta < 20$, the energy is higher than the ground state energy by a very small amount (of the order of $10^{-4}$)
usually undetectable in reasonable calculations. In the inset of Fig.\ref{fig:super} (a zoom of the main figure 
which would correspond to a
typical run for a larger system) we find that the datas are perfectly fitted by 
$R^2(\beta)/(2N)=0.41\beta + 1.24 (1-e^{-\beta/3.4})$ (a usual form) leading to the much smaller 
(and incorrect) $\rho_S\approx 0.41$. 
Hence, we find that it is very important for practical calculations to check both the convergence of the energy,
and the dependence of $\rho_S$ on the GWF very carefully. A poor choice of the GWF will lead to a slow convergence
of the energy (see later in this section) and to incorrect measure of the superfluid density.

\begin{figure}
\includegraphics[width=8cm]{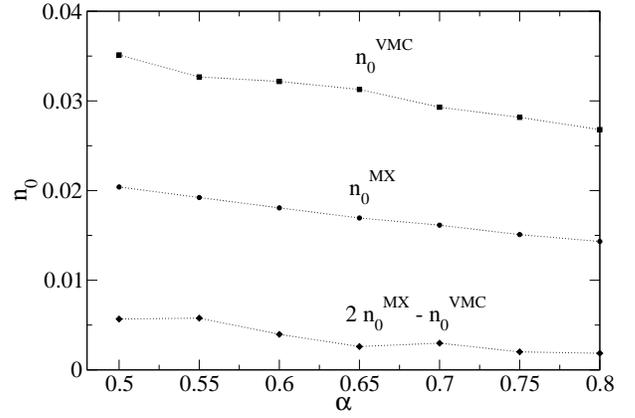}
\caption{\label{conden-frac} Condensate fraction $n_0$ as a function of the variational parameter
$\alpha$, see Eq.~\eqref{trial}.
The three lines correspond to a VMC calculation with $n_0^{\rm VMC}$, a GFMC calculation which gives a mixed 
estimator $n_0^{\rm MX}$ and an extrapolated result  $2 n_0^{\rm MX}-n_0^{\rm VMC}$. This last quantity
is reduced by a factor $3$ between $\alpha=0.5$ and $\alpha=0.8$. }
\end{figure}

A second way to find a supersolid is through non diagonal long range order in the 
one body density matrix
\[
g_1 ({\bf r'},{\bf r}) = \frac{L_x L_y}{N} \moy{c_{\bf r'}^\dagger c_{\bf r}}.
\]
By definition, its largest eigenvalue gives the condensate fraction $n_0$.
For homogeneous systems it leads to $n_0  = (1/ L_x L_y) \sum_{\bf r} g_1 ({\bf r}+{\bf h},{\bf h})$.
Alternatively, $n_0$ is also given by the long-range asymptotic $|{\bf r} - {\bf r'}| \to +\infty$
of $g_1 ( {\bf r'},{\bf r})$ and the two definitions coincide in the thermodynamical limit.
In our case, we reduce significantly finite size effects by excluding the short range part of
$g_1$ with an arbitrary length $r_0$,
\[
n_0  =  \frac{1}{V_{r_0}} \sum_{|{\bf r}| > r_0}
g_1 (  {\bf h} + {\bf r},{\bf h} ),
\]
where $V_{r_0}$ is the system volume excluding a disk of radius $r_0$. 
In contrast to density-density correlations, GFMC only provides
access to mixed estimators for the condensate fraction
since $g_1({\bf r'},{\bf r})$ is a non-local quantity.
Results then depend on the quality of the guiding wavefunction compared to the exact
wavefunction. For small $r_s$, interactions are weak and our variational Bijl-Jastrow 
form~\eqref{guiding} is quite a good approximation. In that case, the bias due to mixed estimators
can be further reduced by computing $2 n_0^{\rm MX}-n_0^{\rm VMC}$ where $n_0^{\rm VMC}$ is the
variational Monte Carlo result using the same guiding wavefunction.
The situation is rather different close to the quantum melting point since the Bijl-Jastrow does not
describe the strong correlations between particles. For this case, the mixed estimator for $g_1$ leads
to results that can not be trusted. To illustrate this point, we plot the condensate fraction
for $r_s = 27$ and $N=50$ particles and for different values of $\alpha$ in Fig.~\ref{conden-frac}. 
The strong dependence on $\alpha$ 
even for  $2 n_0^{\rm MX}-n_0^{\rm VMC}$ clearly rules out the relevance of mixed estimators
to compute condensate fractions. Note that the results presented Fig.~\ref{conden-frac} are quantitatively
only weakly modified after a finite size scaling analysis.
In particular, it is not possible to conclude from these calculations on the presence of a supersolid
with a non vanishing condensate fraction.

\begin{figure}
\vglue +0.05cm
\includegraphics[width=8cm]{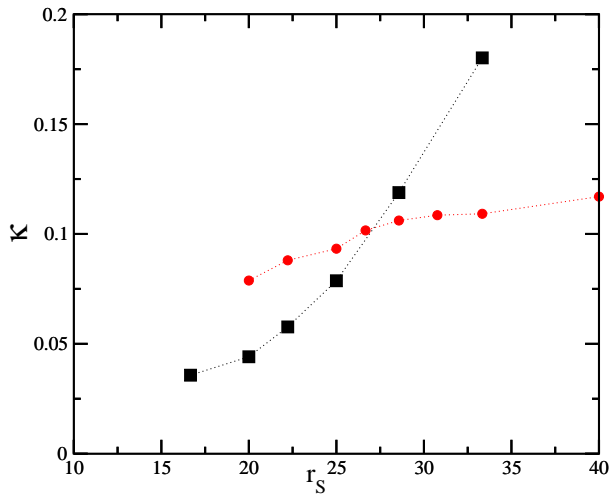}
\caption{\label{fig:over} Parameter $\kappa$ measuring the overlap between the guiding function and the actual 
ground state of the system. $\kappa(r_s)$ at $\alpha=0$ (squares) and $\alpha=0.5$ (circles). The 
datas are taken for $32$ bosons in a  $32\times 56$ grid.  }
\end{figure}

The conclusion of this part is that, although it is not difficult to calculate the energy of the system,
a better GWF is actually needed to get a reliable calculation of superfluid properties. Near the transition point,
our understanding of the physics, and hence the corresponding GWF, is not sufficient. 
The quality of the GWF is usually measured
using either its energy (difference to the ground state) or the variance of its energy (that vanishes for eigenstates).
For practical calculations however, some GWF can have a decent variational energy, but converge very slowly toward
the ground state, hence being poor choice in practice. Following the recent proposal of two of us~\cite{mora2007}, we measure
the quality of the GWF, with its overlap $\Omega$ with the actual ground state of the system. Defining
\be 
\Omega \equiv e^{-\kappa N} \equiv \frac{|\langle\Psi_G|\Psi_0\rangle|^2}
{\langle\Psi_0|\Psi_0\rangle \langle\Psi_G|\Psi_G\rangle}
\ee
the parameter $\kappa$ is given by~\cite{mora2007},
\be
\kappa=\int_0^\infty d\beta\ \  [E(\beta)-E (\beta=+\infty)]
\ee
where $E(\beta)$ is the energy (as a function of imaginary time and per particle) of 
the system while $E (\beta=+\infty)$ is the ground state
energy (per particle). Hence $\kappa$ measures the {\it area} below the curve $E(\beta)$ and is in this sense a good
measure of the practical use of the GWF (a slow converging GFW will have a large $\kappa$ for instance). 
A study of $\kappa$ shown in Fig.~\ref{fig:over} reveals two things: (i) The broken symmetry GWF ($\alpha=0.5$) actually gets better than the BEC ($\alpha=0$) around the transition point $r_s=27$. (ii) However,
in the broken symmetry phase, $\kappa$ still increases with $r_s$, indicating that we describe poorly the crystal
close to the transition. 

\section{Discussion}
\label{sec:conclusions}
\begin{figure}
\vglue +0.2cm
\includegraphics[width=8cm]{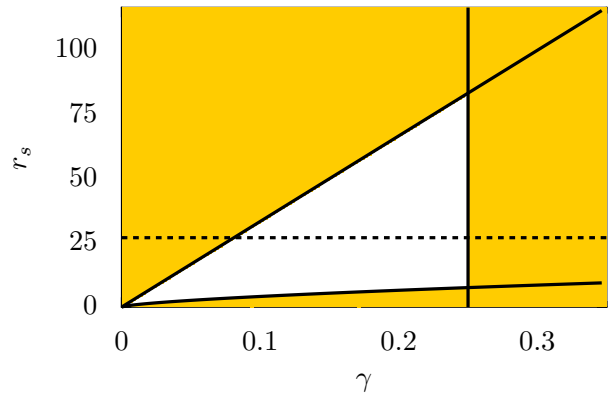}
\caption{\label{fig:exp} Experimental accessible range of parameters in the $(\gamma,r_s)$ plane for a 2D system
of RbCs dipolar molecules. The solid lines indicate the three experimental constraints discussed in the text, while
the dashed line stands for the critical value of $r_s\approx 27$. The white region should be accessible to experiments.}
\end{figure}

We conclude this paper by discussing how the above results would apply to actual experiments
using ultracold dipolar molecular systems. The system we have in mind could be for instance,
RbCs molecules confined by a two-dimensional trap and where electric dipole moments are induced by
a perpendicular electric field. Those types of systems typically have three different experimental
limitations that constrain the range of parameters ($r_s$) than could actually be observed in practice. (i) First, the 
 dipolar forces that can be induced is limited to $C_{\rm dd} = \gamma (e a_0)^2 / (4 \pi \epsilon_0)$ 
($\epsilon_0$ dielectric constant, $a_0$ Bohr radius) 
with $\gamma \le \gamma_{\rm exp}$ (typically $\gamma_{\rm exp}\approx 0.25$ for RbCs~\cite{aymar2005}).
(ii) The maximum density is also limited and $\ell\ge \ell_{\rm exp}$ with a typical value of 
$l_{\rm exp}\approx 60 nm$. It means that $r_s$ is limited by 
$r_s\le (\ell_B^*/\ell_{\rm exp})$. Noting $\ell_B^*=\gamma \ell_0$, we get $r_s\le \gamma(\ell_0/\ell_{\rm exp})$
where $\ell_0\equiv M_0(e a_0)^2 / (\hbar^2 4 \pi \epsilon_0) \approx 20 \mu m$ for RbCs. 
(iii) the temperature is limited to $T_{\rm exp}$ of the order of $T_{\rm exp}\approx 100 nK$. In order to
observe the physics discussed here  $T_{\rm exp}$ should be a fraction $\epsilon$ (say  $\epsilon\approx 1\%$) of 
$E_B$. This imposes, $r_s\ge (T_{\rm exp}/\epsilon T_0)^{1/3} \gamma^{2/3}$ with 
$k T_0=(e a_0)^2 / (8 \pi \epsilon_0\ell_0^3)$. For RbCs, $T_0\approx 1.45 nK$. 
It is an experimental challenge to actually produce ultracold heteronuclear molecules in their 
vibrational and rotational ground state. However we stress here that, once these molecules are produced,
the quantum melting transition itself should be within reach.
Putting these three above conditions 
together, we find the allowed window of parameters shown in  Fig.~\ref{fig:exp}.
 
\acknowledgments
We thank D. Petrov for useful discussions.
C.M. is grateful to the Service de Physique Th\'eorique for hospitality.
C.M also acknowledges support from the DFG-SFB Transregio 12 and the ESF
INSTANS travel program.


\appendix
\section{Ewald summation}\label{appen:ewald}

Finite size effects can be reduced by adding periodic images to the original simulation box forming
a lattice pattern with rectangular boxes of size $L_x\times L_y$. Each atom in the original box
has images in all other boxes. The simple $1/r^3$ dipole interaction between atoms is thus replaced
by an effective interaction taking all images into account, or 
\begin{equation}\label{poten}
V({\bf r}-{\bf r'}) = \sum_{{\bf R}_M} \frac{1}{|{\bf r} - {\bf r'} + {\bf R}_M|^3},
\end{equation}
where ${\bf R}_M=(n L_x,m L_y)$ denote all vectors of a rectangular lattice. Since the summation has 
a poor convergence, we extend the well-known Ewald summation trick to this problem. The $y$-integral in the identity
\begin{equation}\label{form-int}
\frac{1}{|{\bf r}|^3} = \frac{4}{\sqrt{\pi}} \int_0^{+\infty} dy \, y^2 e^{-y^2 |{\bf r}|^2},
\end{equation}
is splitted in two parts  $\int_0^{+\infty} = \int_0^{y_0} + \int_{y_0}^{+\infty}$ 
and $V({\bf r})$ is splitted correspondingly $V({\bf r}) \equiv V^{<}({\bf r}) + V^{>}({\bf r})$.
$y_0$ is a parameter that can be chosen arbitrarily. The second part for $y>y_0$  converges 
exponentially fast when the summation over positions is performed.
One finds
\[
V^{>}({\bf r}) = y_0^3 \, \sum_{ {\bf R}_M} \varphi_{1/2} ( y_0 \, | {\bf r} + {\bf R}_{\rm M}|). 
\]
We have introduced the set of functions
\begin{equation}\label{mishra}
\varphi_{n} (x) = \frac{2}{\sqrt{\pi}} \int_1^{+\infty} dt \, t^n e^{-t x^2},
\end{equation}
which can be expressed as rational functions of ${\rm Erf} (x)$, $e^{-x^2}$ and powers
of $x$. For the case $y<y_0$, we define the function 
$F({\bf r},y) \equiv \sum_{ {\bf R}_M} e^{-y^2  |{\bf r} + {\bf R}_M |^2}$ periodic over the rectangular lattice
for fixed $y$. Thus we write $F({\bf r},y)$ as a summation over its Fourier components on
the reciprocal lattice
\[
F({\bf r},y) = \sum_{{\bf K}_M} e^{i {\bf K}_M \cdot {\bf r} } \frac{\pi}{L_x L_y}
\frac{e^{-{\bf K}_M^2/4 y^2}}{y^2},
\]
with reciprocal lattice vectors ${\bf K}_M=(2 \pi n/ L_x,2 \pi m/ L_y)$,
and the relation
\[
V^{<}({\bf r}) = \frac{4}{\sqrt{\pi}} \int_0^{y_0} d y \, y^2 \, F({\bf r},y),
\] 
yields again an exponentially fast converging expression
for $V^{<}({\bf r})$.  After some straightforward calculations 
and using definition~\eqref{mishra} for functions $\varphi_n(x)$ we finally
arrive at
\begin{equation}\label{fin-inter}
\begin{split}
V({\bf r}) & = y_0^3 \sum_{{\bf R}_M} \varphi_{1/2} ( y_0 \, | {\bf r} + {\bf R}_{\rm M}|) 
\\[3mm]
&+ \frac{\pi y_0}{L_x L_y} \sum_{{\bf K}_M} \cos ( {\bf K}_M \cdot {\bf r} ) \varphi_{-3/2} 
\left( \frac{|{\bf K}_M|}{2 y_0} \right),
\end{split}
\end{equation}
the effective interaction between atoms in the simulation box.
Moreover, subtracting $1/|{\bf r}|^3$ from the right-hand-side of
Eq.~\eqref{fin-inter} and letting ${\bf r} \to {\bf 0}$ 
leads to the constant $\lambda$ in Eq.~\eqref{eq:model},
\[
\begin{split}
\lambda & = - \frac{4}{3 \sqrt{\pi}} y_0^3 + y_0^3 \sum_{{\bf R}_M \ne {\bf 0}} 
\varphi_{1/2} (y_0 \, | {\bf R}_{\rm M}|)  \\[3mm]
&+ \frac{\pi y_0}{L_x L_y} \sum_{{\bf K}_M} \varphi_{-3/2} 
\left( \frac{|{\bf K}_M|}{2 y_0} \right)
\end{split}
\]
which accounts for the self-interaction between each atom and its own images.

The same procedure can be applied to calculate efficiently the kernel~\eqref{kernel} describing
phonons for the crystal phase. Again we start with Eq.~\eqref{form-int}, split the $y$-integral
and $\bar{\Phi} = \bar{\Phi}^{<}+ \bar{\Phi}^{>}$ accordingly. The second part 
converges very rapidly when the summation over the positions ${\bf R}_j$ of the triangular lattice
is performed, one finds 
\[
\begin{split}
\bar{\Phi}_{\alpha,\beta}^{>} & = \frac{1}{8} \sum_{j\ne 0} \left ( \frac{2 R_{\alpha,j} R_{\beta,j}}{b^2} 
\varphi_{5/2} \left( \frac{ |{\bf R}_j|}{b} \right)\right. \\[3mm] & \left.
  - \delta _{\alpha,\beta} 
\varphi_{3/2} \left( \frac{ |{\bf R}_j|}{b} \right) \right) ( 1 - \cos ( {\bf q}\cdot {\bf R}_j) ),
\end{split}
\]
with functions~\eqref{mishra}. For convenience, we choose $y_0=1/b$ where 
$b=\sqrt{2/\sqrt{3}}$ is the triangular lattice spacing.
The first part is computed more efficiently by going to Fourier space in a similar way as above.
The final result reads
\[
\begin{split}
& \bar{\Phi}_{\alpha,\beta}^{<}  = \frac{\pi^3}{2 \sqrt{3}} \sum_{j} \left ( b ({\bf G}_{j}+{\bf q})_\alpha   
b ({\bf G}_{j}+{\bf q})_\beta \times \right. \\[3mm] & \left. \times \varphi_{-3/2} 
\left( \frac{b |{\bf G}_j+{\bf q}|}{2} \right)
  -   b {\bf G}_{\alpha,j}  b {\bf G}_{\beta,j} \,\,
\varphi_{-3/2} \left( \frac{ b |{\bf G}_j|}{2} \right) \right),
\end{split}
\]
where the summation is carried out over all vectors ${\bf G}_j$ on the hexagonal reciprocal lattice with an 
exponentially fast convergence.

\input{boson_crystal6cm.bbl}

\end{document}

%% file: boson_crystal6cm.bbl
\newcommand{{{\PRB}}}{{{Phys. Rev. B}}}\newcommand{{{\PRA}}}{{{Phys. Rev. A}}}\newcommand{{{\PRL}}}{{{Phys. Rev. Lett}}}\newcommand{{{\NPB}}}{{{Nucl. Phys.}}}\newcommand{{{\RMP}}}{{{Rev. Mod. Phys.}}}\newcommand{{{\ADV}}}{{{Adv. Phys.}}}\newcommand{{{\EPJB}}}{{{Eur. Phys. J. B}}}\newcommand{{{\EPJD}}}{{{Eur. Phys. J. D}}}